\documentclass[aps,twocolumn,superscriptaddress,nolongbibliography]{revtex4-1} 
\usepackage[english]{babel}
\usepackage{amsmath, bm, amssymb}
\usepackage{mathtools}
\usepackage{siunitx}
\usepackage{environ}
\usepackage{mathrsfs}
\usepackage{braket}
\usepackage[]{hyperref}
\DeclareMathAlphabet{\mathscrlower}{OT1}{pzc}{m}{it} 
\usepackage{prettyref}
\usepackage{graphicx}
\usepackage{booktabs}
\usepackage{array}
\usepackage{multirow}
\usepackage{dcolumn}
\usepackage{threeparttable}
\usepackage{threeparttablex}
\usepackage{placeins}
\usepackage[stable]{footmisc}
\usepackage{mhchem}
\usepackage{xfrac}
\usepackage{cancel}
\newcommand{\pauli}{\boldsymbol{\sigma}}

\newcommand{\diraccontra}[1]{\boldsymbol{\gamma}^{#1}}

\newcommand{\diraca}{\vec{\boldsymbol{\alpha}}}
\newcommand{\diracg}{\vec{\boldsymbol{\gamma}}}
\newcommand{\diracb}{\boldsymbol{\beta}}

\newcommand{\pos}{\vec{r}}

\newcommand{\momop}{\hat{\vec{p}}}

\newcommand{\efield}{\mathcal{E}}

\newcommand{\Sum}[2]{\sum\limits_{#1}^{#2}}

\newcommand{\braces}[1]{\left\{ #1\right\}}
\let\nablatmp\nabla
\renewcommand{\nabla}{\vec{\nablatmp}}
\DeclarePairedDelimiter\abs{\lvert}{\rvert}
\makeatletter
\let\oldabs\abs
\def\abs{\@ifstar{\oldabs}{\oldabs*}}

\AtBeginDocument{
\heavyrulewidth=.08em
\lightrulewidth=.05em
\cmidrulewidth=.03em
\belowrulesep=.65ex
\belowbottomsep=0pt
\aboverulesep=.4ex
\abovetopsep=0pt
\cmidrulesep=\doublerulesep
\cmidrulekern=.5em
\defaultaddspace=.5em
}
\begin{document}
\title{$\mathcal{CP}$-violation sensitivity of closed-shell radium-containing polyatomic molecular ions}
\date{\today}
\author{Konstantin Gaul}
\email{konstantin.gaul@chemie.uni-marburg.de}
\affiliation{Fachbereich Chemie, Philipps-Universit\"{a}t Marburg, Hans-Meerwein-Stra\ss{}e 4, 35032 Marburg}
\author{Nicholas R. Hutzler}
\author{Phelan Yu}
\affiliation{California Institute of Technology, Pasadena, CA 91125, USA}
\author{Andrew M. Jayich}
\affiliation{Department of Physics, University of California, Santa Barbara, California 93106, USA}
\author{Miroslav Ilia\v{s}}
\affiliation{Department of Chemistry, Faculty of Natural Sciences, Matej Bel University, Tajovského 40, 97401 Banská Bystrica, Slovakia}
\author{Anastasia Borschevsky}
\affiliation{Van Swinderen Institute for Particle Physics and Gravity, University of Groningen, 9747 AG Groningen, The Netherlands}
\begin{abstract}
Closed-shell atoms and molecules such as Hg or TlF provide some of the best low-energy tests of hadronic $\mathcal{CP}$-violation which is considered to be a necessary ingredient to explain the observed excess of matter over antimatter in our universe.   $\mathcal{CP}$-violation is, however, expected to be strongly enhanced in octupole deformed nuclei such as $^{225}$Ra.
Recently, closed-shell radium-containing symmetric-top molecular ions were cooled
sympathetically in a Coulomb crystal [M. Fan \emph{et al.}, Phys. Rev. Lett. 126, 023002 (2021)] and shown to be
well-suited for precision spectroscopy in the search for fundamental physics [P. Yu and N. R. Hutzler, Phys. Rev. Lett. 126, 023003 (2021)]. In closed-shell molecules hadronic $\mathcal{CP}$-violation contributes to a net electric dipole moment (EDM) that violates parity and time-reversal symmetry ($\mathcal{P,T}$), which is the target of measurements. To interpret experiments, it is indispensable to know the electronic structure
enhancement parameters for the various sources of
$\mathcal{P,T}$-violation which contribute to the net $\mathcal{P,T}$-odd EDM. We employ relativistic density functional theory calculations to determine relevant
parameters for interpretation of possible EDM measurements in
\ce{RaOCH3+}, \ce{RaSH+}, \ce{RaCH3+}, \ce{RaCN+}, and \ce{RaNC+} and perform accurate relativistic coupled
cluster calculations of the Schiff moment enhancement in \ce{RaSH+} to gauge the
quality of the density functional theory approach. Finally, we project to bounds on various fundamental $\mathcal{P,T}$-odd
parameters that could be achievable from an experiment with \ce{RaOCH3+} in the near future and asses its complementarity to experiments with Hg and TlF. 
\end{abstract}

\maketitle

\section{Introduction}
Molecules and molecular ions provide some of the best probes of
simultaneous violation of parity and time-reversal symmetry ($\mathcal{P,T}$-violation) \cite{demille:2015}. Within the current experimental resolution a measurement of $\mathcal{P,T}$-violation would be an indirect evidence of
$\mathcal{CP}$-violation beyond the Standard model of particle physics \cite{khriplovich:1997}, which is assumed to be necessary to explain the
imbalance between matter and antimatter (baryon asymmetry) in our universe \cite{sakharov:1967}. Recently, an experiment with
the molecular ion \ce{HfF+} tightened the upper bound on the
electron electric dipole moment (eEDM) \cite{roussy:2023}. Whereas
such experiments with open-shell molecules are established for
searches for $\mathcal{P,T}$-violation in the electron sector,
experiments with closed-shell atoms such as mercury provide some of the best bounds on $\mathcal{P,T}$-violation in the hadronic sector
\cite{graner:2016}. 
%Advantages of closed-shell molecules over atoms shall be exploited in the upcoming %CeNTREX experiment with TlF \cite{grasdijk:2021}. 

In addition to the benefits of relativistic enhancement of $\mathcal{P,T}$-violation in heavy atoms and molecules, the heaviest elements can possess isotopes with octupole deformation, which can significantly enhance the $\mathcal{P,T}$-odd nuclear Schiff moment compared to spherical nuclei such as Tl or Hg
\cite{haxton:1983,auerbach:1996}. In particular, $^{225}$Ra was
identified to posses a large octupole deformation
\cite{butler:1996,butler:2020} and is expected to have a strongly
enhanced nuclear Schiff moment
\cite{spevak:1997,dobaczewski:2005,bishof:2016,dobaczewski:2018}.
Recent developments make precision spectroscopy of molecules containing such short-lived
radioactive isotopes, such as $^{225}$Ra, feasible \cite{garciaruiz:2020,
udrescu:2021,udrescu:2023,wilkins:2023}. 
Moreover, the proposal for direct laser-cooling of
polyatomic molecules \cite{isaev:2016} and its successive realization
\cite{kozyryev:2017:cooling}, were promptly followed by the exploration of advantages of polyatomic molecules for EDM
experiments \cite{isaev:2017,kozyryev:2017:polyedm,anderegg:2023}. Very close-lying $\ell$- or $K$-doublets can serve as internal co-magnetometers and enable large
polarization with weak electric fields, which renders symmetric top and asymmetric top molecules prospective candidates to
search for $\mathcal{P,T}$-violation \cite{kozyryev:2017:polyedm}. 

Radium-containing polyatomic molecular ions combine these advantages of nuclear structure and molecular structure in a single system and can be expected to be exceptionally suitable for tightening bounds on hadronic $\mathcal{CP}$-violation.  Recently, efficient
sympathetic cooling of trapped \ce{RaOH+} and \ce{RaOCH3+} molecules with atomic
radium ions in a Coulomb crystal was demonstrated
\cite{fan:2021}. In this context it was shown that \ce{RaOCH3+} has
favorable properties for fundamental physics experiments
\cite{yu:2021}. Shortly afterwards the asymmetric top molecular ion \ce{RaSH+} was suggested
as a promising candidate for precision experiments as it is assumed to have very close lying $K$-doublets, and subsequently synthesized, trapped and sympathetically cooled \cite{hutzler:2021,jayich:2021,arrowsmithkron:2023}.  

The extraction of bounds on $\mathcal{CP}$-violation from experiments on a fundamental level requires electronic structure theory predictions of the molecular $\mathcal{P,T}$-violation sensitivity coefficients.
In this paper we provide these sensitivity coefficients for the symmetric-top molecules
\ce{RaOCH3+}, \ce{RaCH3+} and the asymmetric-top molecule \ce{RaSH+} and compared to linear Ra-containing molecules \ce{RaNC+} and
\ce{RaCN+}. Relativistic coupled cluster theory is employed
to estimate the influence of electron correlation on the molecular
enhancement of the Schiff moment in \ce{RaSH+}, which is achieved with a new
implementation of the enhancement factor of the Schiff moment
interaction in the DIRAC program package. 
Enhancement of $\mathcal{P,T}$-violation in other molecules and from other possible fundamental sources is computed on the level of density functional theory and subsequently used to project the sensitivity of an experiment with \ce{RaOCH3+} in comparison to experiments with Hg and TlF.

\section{Theory}
\subsection{Effective $\mathcal{P,T}$-odd Hamiltonian}
In a closed-shell asymmetric-top molecule such as \ce{RaSH+} which is
polarized by an external electrical field of strength $\efield$, the
$\mathcal{P,T}$-odd energy shifts are proportional to the interaction
of the nuclear angular momentum, $\vec{I}$, with polarization axis $\vec{\lambda}$ in the molecular frame that can be defined by the principal axes of inertia $\vec{a}$, $\vec{b}$ and $\vec{c}$ as
\begin{equation}
\Delta E_{\cancel{\mathcal{P}},\cancel{\mathcal{T}}}=\vec{I}^\mathsf{T}\cdot
\mathrm{diag}\left(W_{\cancel{\mathcal{P}},\cancel{\mathcal{T}},a}, W_{\cancel{\mathcal{P}},\cancel{\mathcal{T}},b},
W_{\cancel{\mathcal{P}},\cancel{\mathcal{T}},c}\right)\cdot\vec{\lambda}\,,
\end{equation}
where $a,b,c$ denote the components of the vectors in the principal
axes system. In a prolate symmetric-top molecule such as
\ce{RaOCH3+} $b=c$. In analogy to
hyperfine coupling constants, the $b$ and $c$ components contribute to
an anisotropy of the $\mathcal{P,T}$-odd interaction and vanish
in a linear molecule. In the closed-shell molecules \ce{^{225}RaOCH3+}
and \ce{^{225}RaSH+} the electronic structure sensitivity coefficients
$W$ are composed of individual contributions of sources of
$\mathcal{P,T}$-violation as 
\begin{multline}
W_{\cancel{\mathcal{P}},\cancel{\mathcal{T}},a} = d_\mathrm{e} W^\mathrm{m}_{\mathrm{d},a} +
d_\mathrm{sr,n} W_{\mathrm{m},a} + d_\mathrm{sr,n} R_\mathrm{vol} W_{\mathcal{S},a}\\+
\mathcal{S}_\mathrm{coll} W_{\mathcal{S},a} +
k_\mathrm{s} W^\mathrm{m}_{\mathrm{s},a} + k_\mathrm{T}
W_{\mathrm{T},a} + k_\mathrm{p} W_{\mathrm{p},a},
\label{eq: effectHPT}
\end{multline}
where $d_\mathrm{e}$ is electric dipole moment of the electron (eEDM),
$k_\mathrm{s}$ is the scalar-pseudoscalar (SPNEC), $k_\mathrm{T}$ is
the tensor-pseudotensor (TPNEC) and $k_\mathrm{p}$ is the
pseudoscalar-scalar (PSNEC) nucleon-electron current interaction
constant, $d_\mathrm{sr,n}$ is the short-range contribution from quark
EDMs to the neutron EDM (nEDM), $R_\mathrm{vol}$
is nuclear structure enhancement factor for the nEDM contribution to
the Schiff moment and
\begin{equation}
\mathcal{S}_\mathrm{coll}=g\left(a_0\bar{g}^{(0)}_\pi+a_1\bar{g}^{(1)}_\pi+a_2\bar{g}^{(2)}_\pi\right)
\end{equation}
is the collective Schiff moment of $^{225}$Ra from long-range
nucleon-pion interactions, where the strong pion-nucleon
coupling strength is $g\approx13.5$, $\bar{g}^{(0)}_\pi,\bar{g}^{(1)}_\pi,\bar{g}^{(2)}_\pi$ are the nucleon-pion interaction constants and $a_0,a_1,a_2$ are the corresponding nuclear structure enhancement factors (see Ref.~\cite{chupp:2019}). In the following we will drop the subscript $a$ on all $W$ when referring to the principal axis that points approximately along the Ra bonding axes. The individual electronic structure factors $W$ for the heavy Ra nucleus
$A$ are defined as (see
Refs.~\cite{hinds:1980,sushkov:1984,flambaum:1985,martensson-pendrill:1987,flambaum:2002,gaul:2023}):
\begin{widetext}%
\begin{align}
\vec{W}^\mathrm{m}_\mathrm{d}=&  
\frac{\mu_A}{I_A}\left[\Braket{\frac{2c}{e\hbar}\Sum{i=1}{N_\mathrm{elec}}\frac{\imath\diraccontra{0}_i\diraccontra{5}_i}{r_{iA}^3}\hat{\vec{\ell}}_{iA}}
+2\mathrm{Re}\braces{\Sum{j<a}{}\frac{\Braket{j|\frac{2c}{e\hbar}\Sum{i=1}{N_\mathrm{elec}}\imath\diraccontra{0}_i\diraccontra{5}_i\momop^2_i|a}\Braket{a|\frac{\mu_0}{4\pi}\Sum{i=1}{N_\mathrm{elec}}\frac{\pos_{iA}\times\diraca_i}{r_{iA}^3}|j}}{E_j-E_a}}\right]\\ 
\vec{W}_\mathcal{S}          =&  \Braket{\frac{e}{\epsilon_0}\Sum{i=1}{N_\mathrm{elec}}(\nabla_i\rho_A(\pos_i))}   \\
\vec{W}_\mathrm{m}           =&  \eta_A\Braket{4\frac{c\mu_0}{4\pi\hbar}\Sum{i=1}{N_\mathrm{elec}}\frac{\diraca_i}{r_{iA}^{3}}\times\hat{\vec{\ell}}_{iA}}   \\
\vec{W}^\mathrm{m}_\mathrm{s}=&   \frac{\mu_A}{I_A} 2\mathrm{Re}\braces{\Sum{j<a}{}\frac{\Braket{j|\frac{-G_\mathrm{F}Z_A}{\sqrt{2}}\Sum{i=1}{N_\mathrm{elec}}\imath\diraccontra{0}_i\diraccontra{5}_i\rho_A(\pos_i)|a}\Braket{a|\frac{\mu_0}{4\pi}\Sum{i=1}{N_\mathrm{elec}}\frac{\pos_{iA}\times\diraca_i}{r_{iA}^3}|j}}{E_j-E_a}}\\
\vec{W}_\mathrm{T}           =&  \Braket{\sqrt{2}G_\mathrm{F}\Sum{i=1}{N_\mathrm{elec}}\imath\diracg\rho_A(\pos_i)} 
\end{align}
\begin{align}
\vec{W}_\mathrm{p}           =&  \Braket{\frac{G_\mathrm{F}}{\sqrt{2}}\Sum{i=1}{N_\mathrm{elec}}\diracb_i(\nabla_i\rho_A(\pos_i))}  
\end{align}%
\end{widetext}
Here $\Braket{}$ denotes the expectation value for a given
many-electron wave function and $\Ket{j}$, $\Ket{a}$ denote wave
functions of different electronic states with energies $E_j$ and
$E_a$. $c$ is the speed of light, $\hbar$ is the reduced Planck
constant, $\mu_0$ is the magnetic constant, $\epsilon_0$ is the
electric constant, $G_\mathrm{F}$ is the Fermi constant for which we
employ the value \SI{2.22249e-14}{\hartree\bohr\cubed} and
$\mu_\mathrm{N}$ is the nuclear magneton. $\pos_a$,
$\momop_a=-\imath\hbar\nabla_a$ and
$\hat{\vec{\ell}}_{ab}=-\imath\hbar\pos_{ab}\times\nabla_a$ are the
position operator, momentum operator and angular momentum operator of
a particle $a$ relative to particle $b$, respectively. The relative
position of two particles is $\pos_{ab}=\pos_a-\pos_b$ and the
distance operator is $r_{ab}=\left|\pos_{ab}\right|$.  $\rho_A$ is the
normalized nuclear charge density distribution of nucleus $A$. $Z_A$, $I_A$
and $\mu_A$ are the charge number, spin quantum number and magnetic moment of nucleus $A$ respectively, which for \ce{^{225}Ra} are 88, 1/2 and
$-0.7338\,\mu_\mathrm{N}$ \cite{stone:2005} respectively, and
$\eta_A=\frac{\mu_\mathrm{N}}{A_A}+\frac{\mu_A}{A_A-Z_A}$ is
$-0.00091\,\mu_\mathrm{N}$ for \ce{^{225}Ra}. We chose Dirac $\gamma$-matrices to be defined as $\diracg=\begin{pmatrix}\bm{0}&\vec{\pauli}\\-\vec{\pauli}&\bm{0}\end{pmatrix}$,
$\diraccontra{0}=\begin{pmatrix}\bm{1}&\bm{0}\\\bm{0}&-\bm{1}\end{pmatrix}$,
$\diracb=\diraccontra{0}$, $\diraca=\diraccontra{0}\diracg$ and
$\diraccontra{5}=\imath\diraccontra{0}\diraccontra{1}\diraccontra{2}\diraccontra{3}$.

\begin{table*}
\caption{Bond lengths $r$ and bond angles $\phi$ of various Ra-containing
molecular ions with structure \ce{Ra$XY$+} optimized at the level of
DKS-PBE0/dyall.cv3z (DKS) or in case of \ce{RaOCH3+} at the level of SFX2C-CCSD(T)/ANO-RCC-TZVP [CCSD(T)]. In addition to DKS and CCSD(T), structure parameters of \ce{RaSH+} are shown at the level of SFX2C-HF/dyall.cv3z (SFX2C-HF) and SFX2C-PBE0/dyall.cv3z (SFX2C-KS) are shown.}
\label{tab: molstruc}
\begin{tabular}{lll
S[table-format=1.3,round-mode=figures,round-precision=3]
S[table-format=1.3,round-mode=figures,round-precision=3]
S[table-format=3.1,round-mode=figures,round-precision=3]
S[table-format=1.3,round-mode=figures,round-precision=3]
S[table-format=3.1,round-mode=figures,round-precision=3]
S[table-format=3.1,round-mode=figures,round-precision=3]}
\toprule
$X$      & $Y$      && {$r(\ce{Ra-$X$})$/\AA} & {$r(\ce{$X$-$Y$})$/\AA} & {$\phi(\ce{Ra-$X$-$Y$})$/$^\circ$} & {$r(\ce{C-H)})$/\AA} & {$\phi(\ce{$X$-C-H})$/$^\circ$} & {$\phi(\ce{H-C-H})$/$^\circ$}\\
\midrule
S        & H        &SFX2C-HF& 2.871974           & 1.331649            &  98.210 \\
         &          &CCSD(T) & 2.811977           & 1.343828            &  89.874 \\
         &          &SFX2C-KS& 2.799417           & 1.346666            &  93.300 \\
         &          &DKS     & 2.791757           & 1.346366            &  93.843 \\
\\
O        & \ce{CH3} &CCSD(T) & 2.189763           & 1.412486            & 179.994                   & 1.090000         & 110.594               & 108.324 \\
\ce{CH3} & {-}      &DKS     & 2.577462           &                     &                           & 1.097517         & 112.585               & 106.192\\
C        &  N       &DKS     & 2.573846           & 1.159169            & 180.000\\
N        &  C       &DKS     & 2.401855           & 1.179856            & 180.000\\
\bottomrule
\end{tabular}
\end{table*}

\section{Computational Details}
In this work we investigate the contribution from different sources of
$\mathcal{P,T}$-violation on the level of Kohn--Sham (KS) density functional theory (DFT) and
Hartree--Fock (HF) calculations and benchmark our results on accurate relativistic
coupled cluster (CC) calculations of $W_{\mathcal{S}}$. For this purpose we
implemented the integrals of the derivative of the nuclear charge density
distribution in the space of Gaussian basis function $\chi_\mu$ in the
development version of the DIRAC program package \cite{DIRAC23} as 
\begin{equation}
\Braket{\chi_\mu|\nabla \rho_A|\chi_\nu} =
-2\zeta_A\Braket{\chi_\mu|\rho_A\pos_A|\chi_\nu}\,,
\end{equation} 
which are electric dipole moment integrals with a modified
Gaussian density. These integrals are needed for the calculation of
$W_{\mathcal{S}}$ and $W_{\mathrm{p}}$ as operators. The current
implementation was tested by comparison to the results obtained with the program developed in
Ref.~\cite{gaul:2020}.

The Dirac matrix $\imath\diracg$ was not available in the DIRAC program but only the time-reversal anti-symmetric matrix $\diracg$. For 
calculations of $W_\mathrm{T}$ the source code had to be adjusted as described in the appendix.

All relativistic four-component  calculations as well as calculations of
properties in the Levy-Leblond approximation, the exact two-component
approximation (X2C) and its spin-free version (SFX2C) were performed with the
development version of the program package DIRAC \cite{DIRAC23}. 

Dirac--KS (DKS) DFT calculations were performed within the local
density approximation (LDA) using the X$\alpha$ exchange functional
\cite{dirac:1930,slater:1951} and the VWN-5 correlation functional
\cite{vosko:1980}. In comparison to CC calculations DFT functionals tend to underestimate the $\mathcal{P,T}$-odd effects, whereas HF usually overestimates them
\cite{gaul:2017,gaul:2020,gaul:2023}. Therefore, we employed the hybrid LDA
functional with 50\,\% Fock exchange by Becke (BHandH) \cite{becke:1993a}, which was found to give results that agree excellently with CC calculations \cite{gaul:2023}.
Furthermore, we employed the Perdew, Burke , Ernzerhof functional PBE
\cite{perdew:1996} and its hybrid version \cite{adamo:1999}, as well as the
Becke three-parameter hybrid exchange-correlation functional B3LYP
\cite{becke:1993}. To benchmark the exchange-correlation functionals, we
computed the enhancement of the Schiff moment in \ce{RaSH+} on the level of
M\o{}ller-Plesset perturbation theory of second order (MP2) and
single-reference CC with singles and doubles amplitudes (CCSD) and
including perturbative triples [CCSD(T)]. At these correlated levels of
theory, properties are computed in a finite field approach by adding the
perturbing operator with a small amplitude $\lambda$ to the Hamiltonian:
\begin{equation}
\hat{H} =\hat{H}_0+\lambda\hat{H}'\,.
\end{equation}
The energy correction of first order in $\lambda$, $E^{(1)}$ is found by taking the numerical first derivative with respect to $\lambda$:
\begin{equation}
E^{(1)} =\left.\frac{\partial E}{\partial\lambda}\right|_{\lambda=0}\,.
\end{equation}
In all these calculations we computed two points with $\lambda=\pm\SI{1e-9}{}$
for taking the numerical first derivative. The finite field is applied after the DHF mean-field calculation is converged. Therefore, in this approach MP2 results are
orbital unrelaxed (OU-MP2) and, therefore, are expected to be strongly improved
by the CCSD iterations.

Molecular structure parameters of \ce{RaSH+} and \ce{RaOCH3+} were optimized with the
program package CFOUR \cite{cfour,matthews:2020} at the level of scalar relativistic
SFX2C-CCSD(T) calculations employing an atomic natural orbital all
electron basis set of triple zeta quality (ANO-RCC-TZVP) \cite{widmark:1990,veryazov:2004,roos:2004} up to
a change of the geometry gradient of less than
$10^{-5}~E_{\text{h}/a_0}$ as convergence criterion. Wave functions
were optimized until the total energy change between two consecutive HF cycles was less than $10^{-10}~E_{\mathrm{h}}$ or better. The SFX2C-CCSD(T) method provides a good compromise between accuracy and efficiency for the optimization of the molecular structure \cite{chamorro:2022}. The molecular structure of \ce{RaSH+}  optimized in this way was compared to molecular structure optimizations with the program package DIRAC at the level of SFX2C-HF, SFX2C-PBE0 and DKS-PBE0 with the dyall.cv3z basis set \cite{dyall:2009,dyall:2016}. All other molecules were optimized at the level of DKS-PBE0/dyall.cv3z with DIRAC \cite{dyall:2009,dyall:2016}.
The optimized molecular structure parameters are shown in Table
\ref{tab: molstruc}.

In all calculations with DIRAC, the nucleus was described as a
normalized spherical Gaussian nuclear charge density distribution
$\rho_A \left( \vec{r} \right) = \frac{\zeta_A^{3/2}}{\pi ^{3/2}}
\text{e}^{-\zeta_A \left| \vec{r} - \vec{r}_A \right| ^2}$ with
$\zeta_A = \frac{3}{2 r^2 _{\text{nuc},A}}$. The root-mean-square
nuclear charge radius $r_{\text{nuc},A}$ was approximated in
dependence of the nuclear mass number as suggested by Visscher and
Dyall \cite{visscher:1997}. We used the isotopes  $^{1}$H, $^{12}$C, $^{14}$N,
$^{16}$O, $^{32}$S and $^{226}$Ra in all calculations. The influence
of the exact nuclear mass number on properties is negligible for heavy
atoms such as Ra.  
 
\section{Results}
\subsection{Schiff moment enhancement in \ce{RaSH+}}
In the following we estimate the errors of the calculated nuclear Schiff moment
enhancements factors for the molecule \ce{RaSH+} in detail using
sophisticated coupled cluster approaches. This extensive study allows
us to benchmark the DFT calculations, which we successively employ for
other properties and molecules. For all calculations of \ce{RaSH+}
discussed in the following the molecular structure optimized at the
level of SFX2C-CCSD(T)/ANO-RCC-TZVP was used if not stated otherwise.
\subsubsection{Influence of the basis set on the mean-field level}
The influence of the basis set was studied on the level of DHF for
\ce{RaSH+} comparing dyall.cv2z, dyall.cv3z and dyall.cv4z \cite{dyall:2009,dyall:2016}.  From
previous works \cite{gaul:2020,hubert:2022} it is known that the
Schiff moment enhancement operator is extremely sensitive to the wave function
close to the nucleus, and one usually needs to add steep functions with
exponents that are in the range of the exponent of the Gaussian nuclear
charge density distribution. This region is weakly described in the
used Dyall basis sets. Exponential factors $\zeta_i$ of the Gaussian basis functions of form $\mathrm{exp}(-\zeta_i r_i^2)$ of the steepest s and p functions are related by relatively large ratios $\zeta_i/\zeta_j$ of order $\sim4$ and the Gaussians with largest exponential factors are not steep enough to penetrate the nucleus. We added $n$ steep s and $m$ steep p functions with larger exponents, using the ratio of exponents $\zeta_i/\zeta_j$, where $\zeta_i$ and $\zeta_j$ are the largest and second largest exponential factors in the basis set, to obtain the next steeper s and p function with exponent $\zeta_k$ as $\zeta_k=\zeta_i^2/\zeta_j$. This is indicated by ``+$n$s$m$p'' in the descriptor of the basis set. We
also studied the influence of densifying the region of steep exponents
by adding functions with exponential factors obtained as $\zeta_l=\zeta_i\sqrt{\zeta_i/\zeta_j}$ and for all additional $\zeta_k$ analogue. This is indicated in the name of the basis by ``(d)''. Finally, the influence of densifying twice the region of steep functions is studied by adding in addition functions with $\zeta_m=\zeta_i\sqrt[4]{\zeta_i/\zeta_j}$ [indicated by (dd)]. This study is summarized in Table \ref{tab:basisset}. 

\begin{table}[h]
\caption{Influence of the basis set on the enhancement of the nuclear
Schiff moment in \ce{RaSH+} on the level of DHF. In the last column the deviation 
relative to the dyall.cv4z basis set with additional 2s and 2p steep
functions and double densification of the steep function space
[cv4z+2s2p(dd)] is shown.}
\label{tab:basisset}
\begin{tabular}{lllS[round-mode=figures,round-precision=4]S[round-mode=places,round-precision=1]}
\toprule
Basis set & Steeper & Densified &
{$10^{-3}\,W_\mathcal{S}/\frac{e}{4\pi\epsilon_0a_0^4}$} & {Deviation/\%}\\
\midrule
cv2z &      &       & -13.0601&73.5\\
cv3z &      &       & -44.7745& 9.1\\
cv4z &      &       & -46.6597& 5.2\\
cv2z & 1s1p &  no   & -45.6883& 7.2\\ 
cv3z & 1s1p &  no   & -48.1725& 2.2\\
cv2z & 2s2p &  no   & -46.0979& 6.4\\ 
cv3z & 2s2p &  no   & -48.3753& 1.8\\
cv4z & 2s2p &  no   & -48.3142& 1.9\\
cv3z & 3s   &  yes  & -47.3197& 3.9\\
cv2z & 1s1p &  yes  & -49.0113& 0.5\\
cv3z & 1s1p &  yes  & -49.2284& 0.0\\
cv2z & 2s2p &  yes  & -48.8299& 0.8\\
cv3z & 2s2p &  yes  & -49.2247& 0.0\\
cv4z & 2s2p &  yes  & -49.2911& 0.0\\
cv3z & 3s3p &  yes  & -49.2243& 0.0\\
cv3z & 2s2p & double& -49.2449& 0.0\\
cv4z & 2s2p & double& -49.2419&\\
\bottomrule
\end{tabular}
\end{table}
The double densification [cv3z+2s2p(dd), cv4z+2s2p(dd)] has no
significant influence on the results, but we take this augmentation as a reference which describes the region in the nucleus best.
All non-augment basis sets show large deviations to cv4z+2s2p(dd). Even the dyall.cv4z basis results in an error of 5~\% for $W_\mathcal{S}$.
Adding a single steeper function in the s and p block reduces the
error of the cv3z basis already by 7~\%. A densification of the steep
functions finally converges the basis set on the DHF level. Only
adding s functions is not enough to make the error smaller than 1~\%.
In the following calculations we
will use an augmentation with 2s2p and densification, i.e.\ in total
five additional s and p functions [cv3z+2s2p(d)]. 

\subsubsection{Relativistic effects}

To study the influence of relativistic effects we performed
calculations in the non-relativistic limit employing the Levy-Leblond
Hamiltonian, on the scalar relativistic (SFX2C) and two-component
quasi-relativistic level employing the X2C scheme with and without
the atomic mean-field two-electron spin-orbit coupling correction
(AMFI) and on the full
four-component Dirac--Coulomb (DC) and Dirac--Coulomb--Gaunt (DCG)
level. We also computed the effect of explicitly considering the pure
small-component two-electron integrals (SSSS). The results are
summarized in Table \ref{tab:rel}. We find that important relativistic
effects are covered by more than 99\,\% by using a two-component
Hamiltonian, whereas considering only scalar relativistic effects
overestimates the effect by about 30\,\%. The influence of AMFI is
negligible and the corrections due to explicitly considering the full
four-component Hamiltonian almost cancel with the Gaunt correction, which are both
almost 1\,\%. As we are doing coupled cluster calculations in the
following, the used Hamiltonian is not time-critical and we will employ
the DC Hamiltonian for all following discussions. Currently the Gaunt
interaction can not be included in the correlated calculations in
DIRAC.
\begin{table}[h]
\caption{Influence of relativistic effects on the enhancement of the nuclear Schiff moment in
\ce{RaSH+} at the level of DHF/cv3z+2s2p(d).}
\label{tab:rel}
\begin{tabular}{lS[round-mode=figures,round-precision=4]S[round-mode=places,round-precision=1]}
\toprule
Method          & {$10^{-3}\,W_\mathcal{S}/\frac{e}{4\pi\epsilon_0a_0^4}$} &
{Increment/\%}\\
\midrule
Levy-Leblond&  -4.7337 &\\
SFX2C &        -69.8375&+93.2\\
X2C&           -48.8344&+30.1\\
X2C+AMFI&      -48.7911& -0.1\\
%MPZORA&        -49.353\\
DC&            -49.2246& +0.9\\
+G&            -48.8172& -0.8\\
+SSSS&         -48.7239& -0.2 \\
%+QED&          \\
\bottomrule
\end{tabular}
\end{table}

\subsubsection{Electron correlation effects}
The size of the active space is studied on the level of CCSD(T) in Tab
\ref{tab: active_space}. Thereby we were limited to use medium large
active spaces due to the enormous memory requirements. To circumvent
this we studied the influence of larger active spaces with the
cv2z+2s2p(d) basis set.
 
\begin{table}[h]
\caption{Influence of the active space on the RCCSD(T) electron
correlation effects in the enhancement of the nuclear Schiff moment in
\ce{RaSH+}. The first two columns show the lower and upper energy cutoff for active spinors, respectively. In the third column the active space is given as $n$in$m$, where $n$ is the number of electron in $m$ spinors. For the number of spinors a number in parentheses defines the number of spinors for the cv2z+2s2p(d) basis set, whereas the number without parentheses gives the number of spinors for the cv3z+2s2p(d) basis set. The number of digits corresponds to the numerical precision of the finite field CCSD(T) property gradient.}
\label{tab: active_space}
\begin{tabular}{
S[table-format=-2.0]
S[table-format=5.0]
l
S[table-format=-2.2]
S[table-format=-2.2]
S[table-format=-2.3]}
\toprule
\multicolumn{2}{c}{Cutoff/$E_\mathrm{h}$}  & Active space &\multicolumn{3}{c}{$10^{-3}\,W_\mathcal{S}/\frac{e}{4\pi\epsilon_0a_0^4}$} \\
&&&   {cv2z+2s2p(d)} &{cv3z+2s2p(d)} & {cv3z}\\
\midrule
-2&10     &16in300&    &-44.76 \\
-2&30     &16in352& & -44.70 \\
-2&100    &16in420&    &-44.68 \\
-4&30     &26in352(182)&-45.94  &-45.06 & -40.615\\
-4&100    &26in420&    &-45.06  \\
-10&100   &40in420&    &-44.71  \\
{-}&10    &104in(134)    &-45.5 &\\
{-}&30    &104in(182)    &-45.5 &\\
{-}&100   &104in(222)    &-45.5 &\\
%{-}&10000 &104in394&   & \\
\bottomrule   
\end{tabular} 
\end{table}

We find that the active space has no large influence on the results on
the level of CCSD(T). This is in contrast to our findings for the $W_\mathrm{d}$ parameters in BaF, BaOH, YbOH, and BaCH3, where the core electrons have a significant effect on the calculated values \cite{HaaDoeBoe21,DenHaoEli20,ChaBorEli22}. Thus, we choose an active space of
$[-4,30]\,E_\mathrm{h}$ as compromise that allows
computations in a reasonable amount of time. The error of this approximation is
estimated from our calculations to be on the order of $1\,\%$. 

From comparison of the very small cv2z+2s2p(d) basis set with the cv3z+2s2p(d) basis
set for an active space of $[-4,30]\,E_\mathrm{h}$, we see that the
influence of the basis set on the correlation effects on
$W_\mathcal{S}$ is below $2\,\%$. The influence of the steep functions in the basis set is identical to the DHF level: The result with the cv3z basis set deviates by about 10\,\% from the result with the cv3z+2s2p(d) basis set.

\begin{table}[h]
\caption{Influence of electron correlation at the levels of OU-MP2, CCSD,
CCSD(T) and DFT on the enhancement of the nuclear Schiff moment in
\ce{RaSH+}. All calculations are run with an active space of
$[-4,30]\,E_\mathrm{h}$ and the cv3z+2s2p(d) basis set. Deviations
from CCSD(T) calculations are shown in the third column.}
\label{tab: correlation}
\begin{tabular}{lS[round-mode=figures,round-precision=4]S[round-mode=places,round-precision=1]}
\toprule
Method  &  {$10^{-3}\,W_\mathcal{S}/\frac{e}{4\pi\epsilon_0 a_0^4}$} &
{Dev./\%}\\
\midrule
DHF&        -49.2247& 9.2\\
OU-MP2&       -47.9990& 6.5\\
CCSD&      -45.6791& 1.4\\
CCSD(T)&   -45.0646& {-}\\
DKS-BHandH& -43.8859& -2.6\\
DKS-PBE0&   -41.4446& -8.0\\
DKS-B3LYP&  -40.5188& -10.1\\
DKS-PBE&    -37.9984& -15.7\\
DKS-LDA&    -37.1601& -17.5 \\
\bottomrule   
\end{tabular} 
\end{table}

The influence of including electron correlation up to a certain level
was studied by comparing calculations at the levels of DHF, OU-MP2,
CCSD and CCSD(T) in Table \ref{tab: correlation}. Furthermore we
compare to different flavors of density functional theory (DFT). We find that the total correlation effect is only $9\,\%$.
However, using DFT on the level of LDA or GGA the effect of electron
correlation is strongly overestimated leading to too small values and
deviations of more than $15\%$. Including 50\,\% Fock exchange on the
level of LDA (BHandH), the DFT error is reduced and coupled cluster
results can be very well reproduced. This
is in line with previous observations
\cite{gaul:2017,gaul:2020,gaul:2023}. Thus, we will use the BHandH
functional for the calculation of other properties.

The perturbative triples have only a small influence of $1.4\%$ on the results. Based on this, we estimate the overall error due to the higher order
correlations to be below 2\,\%.  

\subsubsection{Molecular structure effects and vibrational corrections}
We compare equilibrium molecular structures for \ce{RaSH+} optimized at different
levels of theory in Table \ref{tab: molstruc}. The influence of
relativistic effects and electron correlation effects on the structure
was determined by comparison of SFX2C-HF and SFX2C-CCSD(T)
calculations as well as by comparing SFX2C-PBE0 and DKS-PBE0
calculations. The former comparison shows that correlation effects are
important, in particular for the bonding angle between Ra, S and H
which is overestimated by HF by about $8\,\%$. The comparison of
SFX2C-PBE0 and DKS-PBE0 calculations shows that spin-orbit coupling
does not play a significant role for the structure of \ce{RaSH+}. The
influence of the molecular structure on the enhancement of the Schiff moment was determined at
the level of DHF/cv3z+2s2p(dd). With the molecular structure computed on the level of SFX2C-HF the value of the
Schiff moment was found to be $-47.75\times10^{3}\,\frac{e}{4\pi\epsilon_0a_0^4}$, which deviates by about $3\,\%$ from the result obtained using the SFX2C-CCSD(T) geometry ($-49.22\times10^{3}\,\frac{e}{4\pi\epsilon_0a_0^4}$).
Thus, the slightly different HF structure has only a minor influence on the
Schiff moment enhancement. From this and the excellent agreement of
molecular structures obtained at the PBE0 and CCSD(T) levels, we expect errors due to the equilibrium molecular structure on the Schiff moment enhancement to be well below $1\,\%$. From considerations of vibrational
corrections to enhancements of $\mathcal{P,T}$-violation in other
polyatomic molecules such as YbOH \cite{gaul:2020a,zakharova:2021} and
RaOH \cite{zakharova:2021a} we can expect that vibrational corrections
are well below $1\,\%$ as well. Therefore,  we are considering molecular structure effects to be negligible at the present level of accuracy.

\subsubsection{Error Budget}
The overall error budget is summarized in Table \ref{tab: errors} and
we estimate the predicted value of $W_\mathrm{S}$ to be accurate
within 7\,\%. Our final value for the Schiff moment enhancement in
\ce{RaSH+} is $-45(3)\times10^{3}\,\frac{e}{4\pi\epsilon_0 a_0^4}$,
which is favorably large and comparable to the enhancement in TlF ($W_\mathrm{S}\sim40\times10^{3}\,\frac{e}{4\pi\epsilon_0
a_0^4}$ \cite{hubert:2022}), although not as large as, for instance, in
multiply charged molecules such as \ce{PaF^3+} or \ce{UF^3+}
\cite{zulch:2022,zulch:2023}. 

\begin{table}
\caption{Estimated uncertainties of the enhancement of the nuclear Schiff moment in
\ce{RaSH+} predicted at the level of CCSD(T)/cv3z+2s2p(d) with active
space $[-4,30]\,E_\mathrm{h}$.}
\label{tab: errors}
\begin{tabular}{p{4cm}S}
\toprule
Error source  &  Amount/\%\\
\midrule
Molecular structure \& vibrational corrections & <1\\
Relativity&  1\\
Basis set&   2\\
Higher order correlation&  2\\
Active space&  1\\
\midrule
Total& <7\\
\bottomrule   
\end{tabular} 
\end{table}

\begin{table*}
\caption{Electronic structure enhancement factors of
$\mathcal{P,T}$-violation in Ra-containing closed-shell molecular ions
computed at the level of DCKS-BHandH/dyall.cv3z+sp(d).}
\label{tab: all_results}
\begin{tabular}{
l
S[table-format = -2.1, scientific-notation=fixed,fixed-exponent=0,round-mode=figures, round-precision=3]
S[table-format = -3.0, scientific-notation=fixed,fixed-exponent=0,round-mode=figures, round-precision=3]
S[table-format = -1.2, scientific-notation=fixed,fixed-exponent=0,round-mode=figures, round-precision=3]
S[table-format = -2.1, scientific-notation=fixed,fixed-exponent=0,round-mode=figures, round-precision=3]
S[table-format = -1.2, scientific-notation=fixed,fixed-exponent=0,round-mode=figures, round-precision=3]
S[table-format = -1.2, scientific-notation=fixed,fixed-exponent=0,round-mode=figures, round-precision=3]
}
\toprule
Molecule & 
\multicolumn{1}{c}{$W^\mathrm{m}_\mathrm{d}/\frac{10^{20}\,\mathrm{Hz}\,h}{e\,\mathrm{cm}}$} &
\multicolumn{1}{c}{$W^\mathrm{m}_\mathrm{s}/(h\,\mathrm{Hz})$} &
\multicolumn{1}{c}{$W_\mathrm{T}/(h\,\mathrm{kHz})$} &
\multicolumn{1}{c}{$W_\mathrm{p}/(h\,\mathrm{Hz})$}  &
\multicolumn{1}{c}{$W_\mathrm{m}/\frac{10^{17}\,\mathrm{Hz}\,h}{e\,\mathrm{cm}}$} &
\multicolumn{1}{c}{$W_\mathcal{S}/\frac{\mathrm{MHz}\,h}{e\,\mathrm{fm}^3}$} \\
\midrule 
\ce{RaSH+}   & 31.92107 & 82.8961 & -3.91312 & -15.3459 & -1.68312 & -1.94862 \\
\ce{RaOCH3+} & 34.85017 & 93.4798 & -4.4495 & -17.4597 & -1.87528 & -2.23383 \\
\ce{RaCH3+}  & 38.98652 & 98.4958 & -4.61523 & -18.0855 & -1.91484 & -2.24067 \\
\ce{RaCN+}   & 32.53858 & 86.373 & -4.09946 & -16.0856 & -1.82037 & -2.05838 \\
\ce{RaNC+}   & 32.03558 & 86.1227 & -4.1006 & -16.0975 & -1.8211 & -2.07582 \\
\bottomrule
\end{tabular}
\end{table*}

\subsection{$\mathcal{P,T}$-odd electronic structure enhancement
factors and projected limits}
At the level of DKS-BHandH, we computed properties that are considered to be 
relevant for $\mathcal{P,T}$-violation in closed shell molecules [see
eq. (\ref{eq: effectHPT})] for \ce{RaSH+}, \ce{RaOCH3+}, \ce{RaCH3+}, \ce{RaNC+}
and \ce{RaCN+}. The results are shown in Tab. \ref{tab: all_results}.
We find for all compounds very similar enhancement effects, and the
influence of the substituents is small. 

\subsubsection{Anisotropy and asymmetry of $\mathcal{P,T}$-odd
coupling tensors in asymmetric-top molecules}
As outlined in the theory section, in an asymmetric-top molecule the
electronic enhancement factors are rank-1 tensors in
principle. As the moment of inertia along the Ra--S bond in \ce{RaSH+} is very small, coupling tensors will be similar to those of a linear molecule. Thus,  no considerable
enhancement effects along the principal axes perpendicular to the Ra--S bond can
be expected. However, it is interesting to consider if the asymmetry is
different for different $\mathcal{P,T}$-odd properties. This could
be utilized in other system with larger difference in the principal axes to disentangle different contributions by measurements
with different polarization directions. Though achieving controlled polarization in only one direction is already difficult for a polyatomic molecule, the different principal axes have been manipulated and studied with alternating fields for chiral spectroscopy~\cite{patterson:2013}, and such techniques could potentially be combined with measurement schemes which use alternating fields for $\mathcal{P,T}$-violation searches~\cite{verma:2020,takahashi:2023,zhang:2023}

In analogy to a diagonal rank-2 tensor we can define the
isotropy $W_{\cancel{\mathcal{P}},\cancel{\mathcal{T}},\mathrm{iso}}$, the anisotropy
$W_{\cancel{\mathcal{P}},\cancel{\mathcal{T}},\mathrm{ani}}$ and the asymmetry $W_{\cancel{\mathcal{P}},\cancel{\mathcal{T}},\mathrm{asy}}$ as
\begin{align}
W_{\cancel{\mathcal{P}},\cancel{\mathcal{T}},\mathrm{iso}} &= \frac{W_{\cancel{\mathcal{P}},\cancel{\mathcal{T}},a}+
W_{\cancel{\mathcal{P}},\cancel{\mathcal{T}},b}+ W_{\cancel{\mathcal{P}},\cancel{\mathcal{T}},c}}{3} \\
W_{\cancel{\mathcal{P}},\cancel{\mathcal{T}},\mathrm{ani}} &=
\frac{2W_{\cancel{\mathcal{P}},\cancel{\mathcal{T}},a}-W_{\cancel{\mathcal{P}},\cancel{\mathcal{T}},b}-W_{\cancel{\mathcal{P}},\cancel{\mathcal{T}},c}}{3} \\
W_{\cancel{\mathcal{P}},\cancel{\mathcal{T}},\mathrm{asy}}
&=(W_{\cancel{\mathcal{P}},\cancel{\mathcal{T}},b}-W_{\cancel{\mathcal{P}},\cancel{\mathcal{T}},c})/W_{\cancel{\mathcal{P}},\cancel{\mathcal{T}},\mathrm{ani}} 
\end{align}
For \ce{RaSH+} the isotropy and anisotropy are in good approximation
proportional to $W_{\cancel{\mathcal{P}},\cancel{\mathcal{T}},a}$, as $W_{\cancel{\mathcal{P}},\cancel{\mathcal{T}},a}$ is expected to be much
larger than the other two components. Thus, we focus on the asymmetry. We computed the $b$ and $c$ components of $W_{\mathcal{S}}$ and $W_{\mathrm{d}}$ for \ce{RaSH+} on the level
of DKS-BHandH. The $c$ components are zero within the numerical precision and for the $b$  components we find $W_{\mathcal{S},b}=2\,\frac{\mathrm{kHz}\,h}{e\,\mathrm{fm}^3}$, $W_{\mathrm{d},b}=0.1\,\frac{10^{20}\,\mathrm{Hz}\,h}{e\,\mathrm{cm}}$. As expected the resulting asymmetry is negligibly small
($W_{\mathcal{S},\mathrm{asy}}\sim-2\times10^{-3}$, $W_{\mathrm{d,asy}}\sim 6\times10^{-3}$). Interestingly, the asymmetry for the eEDM enhancement differs considerably from the Schiff moment enhancement asymmetry, which
reflects different ratios of $W_\mathcal{S}/W_\mathrm{d}$ for different polarization directions. This could be valuable for the disentanglement of the different sources of
$\mathcal{P,T}$-violation \cite{jung:2013,chupp:2015,fleig:2018,gaul:2019,gaul:2023} in an experiment with an asymmetric top molecule, if polarization in different directions could be controlled. For \ce{RaSH+}, effects are anyway on the order of $10^{-3}$ and, therewith, too small to be relevant. It would be
interesting to exploit larger differences between the principal components of $\mathcal{P,T}$-odd enhancement factors,
which can be expected in heavy-element-containing chiral
molecules, in which other parity-violating forces beyond the Standard
model of particle physics are favorably enhanced
\cite{gaul:2020c,gaul:2020d}.

\begin{table*}
\begin{threeparttable}
\caption{Projected limits on fundamental sources of
$\mathcal{P,T}$-violation from a proposed experiment with single \ce{RaOCH3+} molecular ion 
compared to the Hg \cite{graner:2016} and TlF \cite{cho:1991} experiments. Limits are derived from
the projected experimental sensitivity of an experiment with
\ce{RaOCH3+} $\delta\nu\approx\SI{6.5e-5}{\hertz}$ \cite{yu:2021}, the electronic
structure enhancement factors given in Table \ref{tab: all_results}
and the nuclear structure factors discussed in the text.}
\label{tab: proj_limits}
\begin{tabular}{c
S[table-format=1.0e-2]
S[table-format=1.0e-2]
S[table-format=1.0e-2]
S[table-format=1.0e-2]
S[table-format=1.0e-2]
S[table-format=1.0e-2]
S[table-format=1.0e-2]
S[table-format=1.0e-2]
S[table-format=1.0e-2]
}
\toprule
System & {$d_\mathrm{e}/(\si{\elementarycharge\centi\meter})$} &
{$d_\mathrm{sr,n}/(\si{\elementarycharge\centi\meter})$} &
{$d_\mathrm{sr,p}/(\si{\elementarycharge\centi\meter})$} &
{$k_\mathrm{s}$} &
{$k_\mathrm{T}$} & 
{$k_\mathrm{p}$} &
{$g_\pi^{(0)}$}& {$g_\pi^{(1)}$}& {$g_\pi^{(2)}$}\\
\midrule
TlF\tnote{a}&6e-26&{-}&6e-24&2e-6&6e-8&2e-5&2e-8&9e-10&6e-9\\
Hg\tnote{a}&3e-27&2e-26&{-}&2e-8&2e-10&4e-8&3e-11&2e-11&2e-11\\
\midrule
\ce{RaOCH3+} & 4e-26 & 2e-24 &{-} & 2e-6 & 3e-8 & 8e-6& 4e-11 & 1e-11 & 2e-11\\
\bottomrule
\end{tabular}
\begin{tablenotes}
\item[a] Values are computed with the experimental uncertainty on the
EDM of TlF $\sigma_d=\SI{2.9e-23}{\elementarycharge\centi\meter}$ with
an external electric field for polarization of strength
$\mathcal{E}=\SI{16000}{\volt\per\centi\meter}$ from
Ref.~\cite{cho:1991} and the EDM of Hg
$\sigma_d=\SI{3.1e-30}{\elementarycharge\centi\meter}$ from
Ref.~\cite{graner:2016}. Electronic structure enhancement factors $W$ for Hg and TlF were computed at the
level of ZORA-BHandH in Ref.~\cite{gaul:2023}. Nuclear structure
factors are taken from Ref.~\cite{chupp:2019}.
\end{tablenotes}
\end{threeparttable}
\end{table*}

\begin{figure*}
\includegraphics[width=.925\textwidth]{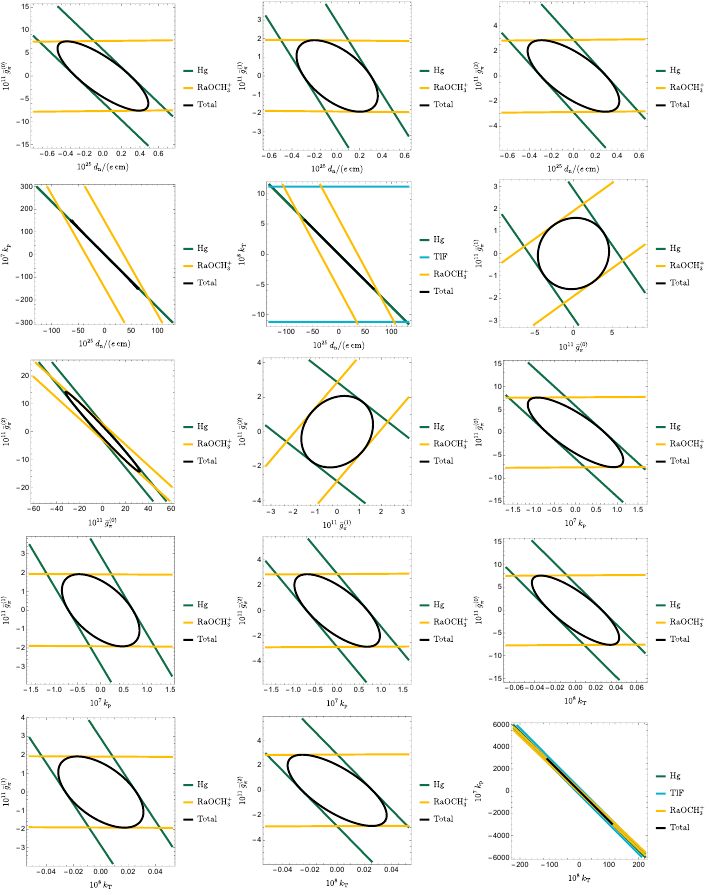}
\caption{Restriction of two dimensional subspaces including parameters 
$d_\mathrm{sr,n}$, $k_\mathrm{T}$, $k_\mathrm{p}$, $g_\pi^{(0)}$, $g_\pi^{(1)}$, $g_\pi^{(2)}$ of the considered nine dimensional space of $P,T$-odd parameters by experiments with Hg, TlF and the proposed experiment with \ce{RaOCH3+}. Coverage regions are computed with electronic structure parameters of \ce{RaOCH3+} provided in this work,
electronic structure parameters of \ce{TlF} and \ce{Hg} from
Ref.~\cite{gaul:2023} and nuclear structure parameters from
Ref.~\cite{chupp:2019}. Experimental uncertainty on the
EDM of TlF $\sigma_d=\SI{2.9e-23}{\elementarycharge\centi\meter}$ with
an external electric field for polarization of strength
$\mathcal{E}=\SI{16000}{\volt\per\centi\meter}$ is taken from
Ref.~\cite{cho:1991} and on the EDM of Hg
$\sigma_d=\SI{3.1e-30}{\elementarycharge\centi\meter}$ is taken from
Ref.~\cite{graner:2016}. The expected uncertainty of an experiment with a single 
\ce{RaOCH3+} molecule $\delta\nu\approx\SI{6.5e-5}{\hertz}$ is used as proposed
in Ref.~\cite{yu:2021}. All bounds are computed with Gaussian
probability distributions of 95\,\%\ CL as described in
Ref.~\cite{gaul:2023}. The two dimensional subspaces that include the parameters $d_\mathrm{e}$, $d_\mathrm{sr,p}$, $k_\mathrm{s}$ are shown in the supplementary material.}
\label{fig: globalbounds}
\end{figure*}

\subsubsection{Projected limits on fundamental sources of
$\mathcal{P,T}$-violation}
Ref.~\cite{yu:2021} gave the estimated experimental
sensitivity for precession frequency measurements on a single trapped
\ce{^{225}RaOCH3+} ion with two weeks of data taking to be
$\delta
\nu=\SI{7.5}{mrad\per\second\per\sqrt{336}}/2\pi=\SI{6.5e-5}{\hertz}$.
Assuming this sensitivity, we can project limits on fundamental sources
of $\mathcal{P,T}$-violation from an experiment with
\ce{^{225}RaOCH3+} within single-source models, i.e.\ assuming only a
single source of $\mathcal{P,T}$-violation being existent at a time. 

For the collective Schiff moment of \ce{^{225}Ra} we use recommended values from
Ref.~\cite{chupp:2019}:

\begin{equation}
a_0=\SI{-1.5}{\elementarycharge\femto\meter\cubed};\,
a_1=\SI{6}{\elementarycharge\femto\meter\cubed};\,
a_2=\SI{-4}{\elementarycharge\femto\meter\cubed}\,.
\end{equation}

To our best knowledge there is no calculation for $R_\mathrm{vol}$
available.  Using a rough estimate as discussed in
Ref.~\cite{ginges:2004} we obtain
$R_\mathrm{vol}\sim\frac{1}{10}(\SI{1.2}{\femto\meter})^2
\frac{3}{5}A^{2/3}=\frac{1}{10}(\SI{1.2}{\femto\meter})^2
\frac{3}{5}\times225^{2/3}=\SI{3.2}{\femto\meter\squared}$

In Table \ref{tab: proj_limits} we compare
the resulting projected sensitivities for \ce{RaOCH3+} with the
single-source limits of the Hg \cite{graner:2016} and TlF \cite{cho:1991} experiments, which we computed from the
electronic structure data provided in Ref.~\cite{gaul:2023} and
nuclear structure data summarized in Ref.~\cite{chupp:2019}. We want to note that \ce{RaOCH3+} and \ce{Hg} are sensitive to the nEDM whereas TlF is sensitive to the short-range proton EDM (pEDM) $d_\mathrm{sr,p}$. Moreover, open-shell molecules such as \ce{HfF+}, \ce{ThO} or RaF are generally much more sensitive to $d_\mathrm{e}$ and $k_\mathrm{s}$, which we will, therefore, not discuss in the following.

We find that an experiment with even a single \ce{RaOCH3+} molecule will have sensitivity to $\mathcal{P,T}$-violation comparable to that of the Hg
experiment for certain underlying sources. In particular, the single-molecule experiment with \ce{RaOCH3+} would place similar bounds
on pion-nucleon couplings $\bar{g}_{\pi}^{(0)}$, $\bar{g}_{\pi}^{(1)}$, $\bar{g}_{\pi}^{(2)}$ and, therewith, similar bounds on the
$\mathcal{CP}$-violation parameter $\bar{\theta}\sim
\bar{g}_{\pi}^{(0)}/0.015$ \cite{vries:2015} and quark EDMs
$\bar{d}_\mathrm{d}-\bar{d}_\mathrm{u}\sim\SI{5e-15}{\elementarycharge\centi\meter}\,\bar{g}_\pi^{(1)}$ \cite{chupp:2019}.  In comparison to the TlF experiment of
Ref.~\cite{cho:1991}, only a slightly increased sensitivity can be expected
for the electron-nucleon current
interactions, whereas sensitivity to pion-nucleon couplings is
increased by at least two orders of magnitude. The sensitivity of the TlF experiment is expected to be improved by roughly three orders of magnitude in the upcoming CeNTREX experiment with TlF, for which a frequency shift of $\delta\nu=\SI{50e-9}{\hertz}$ is claimed to be achievable \cite{grasdijk:2021}. Taking this projected sensitivity for TlF would surpass all single-source bounds from Hg and an experiment with a single \ce{RaOCH3+}. Having a valence proton and therewith sensitivity to $d_\mathrm{sr,p}$ instead to $d_\mathrm{sr,n}$, TlF would at any rate be complementary to \ce{RaOCH3+}.

The limits here are given for a single trapped molecule with a coherence time of $\tau=5$~s limited by blackbody pumping at 300~K~\cite{yu:2021}.  Increasing the coherence time by a factor of $t$, which would require using a cryogenic apparatus, and trapping $N>1$ ions at a time~\cite{roussy:2023,ng:2022,zhou:2023}, would increase the sensitivity by a factor of $\sqrt{tN}$, as would using quantum-enhanced metrology to beat the standard quantum limit~\cite{zhang:2023}.  Experimental efforts to realize these advances are currently underway.

The single-source assumption is not a realistic model
\cite{jung:2013,chupp:2015,gaul:2023} and in a global model for $\mathcal{P,T}$-violation the
polyatomic Ra-containing molecular ions will be very beneficial as
ratios between different sensitivity coefficients vary considerably
compared to Hg and TlF. The complementarity of \ce{RaOCH3+} to Hg and
TlF is illustrated for the two-dimensional subspaces of the considered
$\mathcal{P,T}$-odd parameter space in Fig.~\ref{fig: globalbounds} as suggested in Ref.~\cite{gaul:2023}. The subspaces which include the parameters $d_\mathrm{e}$, $d_\mathrm{sr,p}$, $k_\mathrm{s}$ to which the \ce{RaOCH3+} molecule is rather insensitive compared to open-shell systems and systems with a valence proton are shown in the supplementary material. Fig.~\ref{fig: globalbounds} highlights the complementary sensitivity to pion-nucleon interactions of Ra-containing polyatomic ions compared to the Hg experiment.

\section{Conclusion}
We have computed electronic structure enhancement of
$\mathcal{P,T}$-violation in radium containing molecular ions with closed electronic shells, which
are promising candidates for precision tests of fundamental symmetry violation in the hadronic sector. With a new implementation of the Schiff moment enhancement operator in DIRAC we could perform accurate
coupled cluster calculations to gauge the importance of electron
correlation for this effect in the asymmetric-top molecule \ce{RaSH+}. The suggested value for the Schiff moment enhancement in
\ce{RaSH+} is $-45(3)\times10^{3}\,\frac{e}{4\pi\epsilon_0 a_0^4}$.
Subsequently we used this approach to benchmark DFT functionals for the computation
of enhancement factors for various possible sources of
$\mathcal{P,T}$-violation in several polyatomic radium-containing
molecular ions. The asymmetry of
$\mathcal{P,T}$-violation enhancement in \ce{RaSH+} was studied within DFT. Although this effect is too small to be relevant in \ce{RaSH+}, our findings indicate that asymmetric-top molecules which deviate strongly from prolate or oblate rotors may provide an interesting route for disentanglement of sources of $\mathcal{P,T}$-violation via measurements with different polarization directions.
Finally, we computed projected $\mathcal{P,T}$-violation sensitivities of an experiment with a single \ce{RaOCH3+} molecule. Our results show that enhancements in radium-containing
polyatomic molecular ions are favorably large and are complementary to those in the TlF and Hg experiments. 

\begin{acknowledgments}
K.G. is indebted to Robert Berger for support and acknowledges the Deutsche
Forschungsgemeinschaft (DFG, German Research Foundation) --- Projektnummer 328961117 --- SFB 1319 ELCH and the Haeuser-Stiftung for funding a research stay in Groningen. K.G. thanks Pi A. B. Haase, Yuly M.  Comorro Mena and I.  Agust\'in Aucar for discussions. A.M.J acknowledges support from the US Department of Energy (DE-SC0022034).  N.R.H acknowledges support from NSF CAREER Award No. PHY-1847550. Computer time on the Peregrine high performance computing cluster provided by the Center for Information Technology
of the University of Groningen is gratefully acknowledged.  
\end{acknowledgments}

\appendix
\section{Implementation of $W_\mathrm{T}$}
Each matrix in the vector $\imath\diracb\diraca$ can be decomposed in a quaternion
scalar and a \emph{real valued} matrix in the space of the large and
small component spinors:
\begin{equation}
\imath\diracb\diraca=-\imath \pauli_y \otimes\imath\vec{\pauli} \equiv
\begin{pmatrix}\hat{k}\\\hat{j}\\\hat{i}\end{pmatrix} \otimes(-\imath \pauli_y) =
\begin{pmatrix}\hat{k}\\\hat{j}\\\hat{i}\end{pmatrix}\otimes
\begin{pmatrix}0&-1\\1&0\end{pmatrix}\,,
\end{equation}
where $\hat{i},\hat{j},\hat{k}$, with $\hat{i}=\imath$ are the quaternion units that correspond to the space spanned by $\imath\vec{\pauli}$ with $\imath=\sqrt{-1}$ being the imaginary unit.
This matrix is time-reversal symmetric and can be combined with
anti-symmetric real valued basis-function integrals to form hermitian operators. In the DIRAC program the
matrix defined under this name \texttt{iBETAAL} ($\imath\diracb\diraca$) is a time-reversal anti-symmetric matrix of the form:
\begin{equation}
\diracb\diraca=-\imath^2\pauli_y \otimes\imath\vec{\pauli} \equiv
\imath\begin{pmatrix}\hat{k}\\\hat{j}\\\hat{i}\end{pmatrix} \otimes(-\imath \pauli_y) =
\begin{pmatrix}-\hat{j}\\\hat{k}\\1\end{pmatrix}\otimes
\begin{pmatrix}0&-1\\1&0\end{pmatrix}\,.
\end{equation}
For the computation of 
$W_\mathrm{T}$ the definition of \texttt{iBETAAL} in DIRAC was
modified to
\begin{equation}
\begin{pmatrix}\hat{k}\\\hat{j}\\\hat{i}\end{pmatrix} \otimes \begin{pmatrix}0&-1\\1&0\end{pmatrix}\,
\end{equation}
a time-reversal symmetric matrix which is combined with the
anti-symmetric real-valued integrals of the normalized nuclear charge
density distribution operator $\rho_A$.

\bibliography{ra_poly_ions_cpv.bib}

\end{document}